\documentclass[preprint]{aastex}

\pdfoutput=1

\shorttitle{Aircraft Illumination Avoidance}
\shortauthors{Coles et al.}

\begin{document}

\title{A Radio System for Avoiding Illuminating Aircraft with a Laser
Beam}

\author{W.\,A. Coles, T.\,W. Murphy Jr., J.\,F. Melser, J.\,K. Tu, G.\,A.
White, K.\,H. Kassabian, K. Bales and B.\,B. Baumgartner}
\affil{University of California, San Diego, 9500 Gilman Drive, La Jolla, CA
92093}
\email{bcoles@ucsd.edu, tmurphy@physics.ucsd.edu}

\begin{abstract}
When scientific experiments require transmission of powerful
laser or radio beams through the atmosphere the Federal Aviation
Administration (FAA) requires that precautions be taken to avoid
inadvertent illumination of aircraft. Here we describe a highly reliable
system for detecting aircraft entering the vicinity of a laser beam by making
use of the Air Traffic Control (ATC) transponders required on most
aircraft. This system uses two antennas, both aligned with the laser beam.
One antenna has a broad beam and the other has a narrow beam. The ratio of
the transponder power received in the narrow beam to that received in the
broad beam gives a measure of the angular distance of the aircraft from the
axis that is independent of the range or the transmitter power. This ratio
is easily measured and can be used to shutter the laser when the aircraft
is too close to the beam.  Prototype systems operating on astronomical
telescopes have produced good results.
\end{abstract}


\section{Introduction}

A number of scientific experiments require the transmission of a laser beam
through the atmosphere, using an astronomical telescope or its equivalent.
In order to avoid hazard to aircraft the FAA requires that one or more
observers be stationed outside any telescope that is transmitting a laser
beam. These observers close the laser shutter when an aircraft is observed
within $25^\circ$ of the laser beam (as viewed from the telescope). Such
experiments include: lunar and satellite laser ranging \citep{llr,ilrs};
creation of artificial guide stars for active optics
\citep[e.g.,][]{keck-lgs}; and atmospheric remote sensing using lidar
\citep{lidar1,lidar2}. In this paper we discuss part of an aircraft
detection system now employed at the Apache Point Observatory (APO) for a
lunar ranging experiment called APOLLO \citep{apollo}. This system could be
used not only for the other laser beam experiments mentioned above, but
also for protecting aircraft from high-powered radar transmitters such as
those used for ionospheric research \citep[e.g.,][]{ionosphere}.  The
detection scheme described here is used in conjunction with a complementary
infrared camera detection system, together providing robust protection to
aircraft.

The FAA rules effectively require transponders on all commercial aircraft
and most private aircraft \citep[the exact language can be found in Section
91.215 of the Federal Aviation Regulations:][]{far}.  These transponders
are interrogated frequently (at 1030~MHz) by the regional ATC radars and
also by the airborne Traffic Collision Avoidance System (TCAS). The
transponders reply incoherently at $1090\pm 3$~MHz with a pulse coded
response. The response must have vertical electric field polarization, an
omni-directional pattern, and transmitted peak power between 70 and 500~W.
Various coding schemes convey information about the aircraft.  Mode-A and
Mode-C responses communicate a temporarily assigned aircraft identity and
altitude, respectively.  A newer Mode-S encoding flexibly communicates
permanent aircraft identity, coordinates, altitude, etc., but still as
pulsed transmission at 1090~MHz.

\begin{figure}
\begin{center}\includegraphics[width=89mm, angle=0]{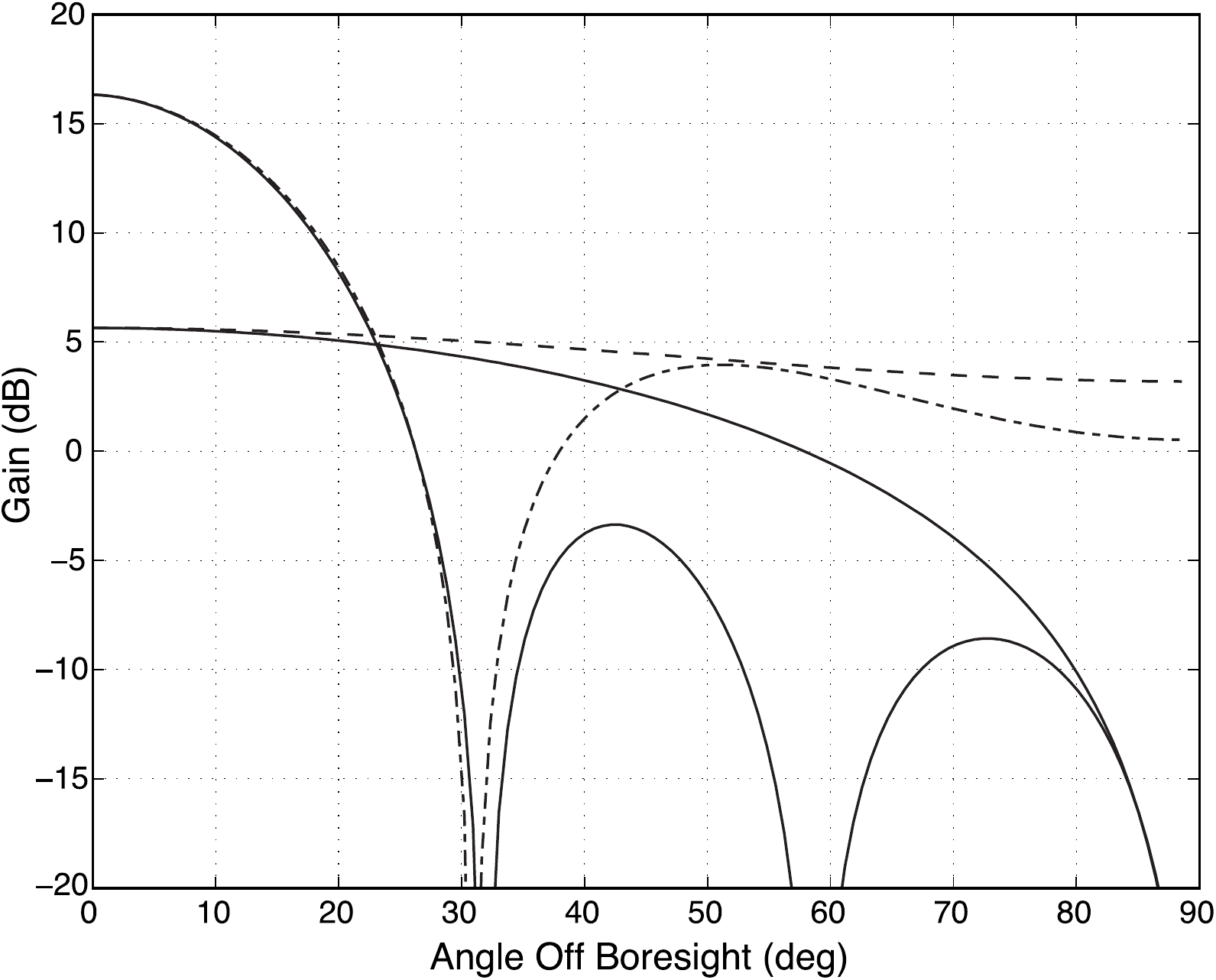}
\end{center}

\caption{Gain of the broad-beam and narrow-beam antennas as a function of
the angular separation of the transponder from the axis. The solid lines 
are H-plane cuts through the gains of the antennas actually
used and the dashed lines are E-plane cuts. If the antennas
are pointed to zero elevation, the E plane is the vertical plane and the H
plane is horizontal.  The power received is proportional to the gain.
\label{fig:gains}}

\end{figure}

The APOLLO laser is never used at elevations less than $15^\circ$ and this
elevation restriction is typical of other laser and radar transmitters.
Thus for practical altitudes of $< 13$~km the aircraft range will not
exceed $\sim$50~km, and the received power is very high by modern
communications standards ($> -69$~dBm) so that it may be easily detected
with a total power receiver. However the received power is highly variable
because both the range and the transmitted power are variable.  The design
requirement is a highly reliable method of detecting when an aircraft
transponder is within about $15^\circ$ of the telescope beam. This
$15^\circ$ specification---differing from the 25$^\circ$ angle used by
human spotters---is set by the expected angular rate of aircraft,
transponder interrogation frequency, and the desire to avoid excessive
triggers when pointing the beam as low as $15^\circ$ above the horizon.
The general concept is to use two antennas aligned with the optical axis of
the telescope, one with a beam width (full-width at half-power) of about
$30^\circ$ and the other with a beam width of about $90^\circ$, as shown in
Figure~\ref{fig:gains}. The ratio of the power received by the narrow beam
antenna to that received by the broad beam antenna depends only on the
angular position of the transponder with respect to the beam axis. In
particular it does not depend on the distance, transmitted power, or
polarization mismatch.

The APO telescope, like many others used in this type of experiment, is on
an altazimuth mount with a secondary mirror more than 80~cm in diameter
centered on the optical axis.  The radio antennas can be mounted facing the
sky on the secondary support structure and aligned with the optical axis
without interfering with the optical beam, and the polarization will remain
vertical as the telescope is moved. This geometry is also typical of the
radar antennas used for ionospheric research although the secondary
reflector is much larger in these cases.  For telescopes on an equatorial
mount the position angle of the linear polarization changes with hour
angle. This variation can be easily accommodated by changing to antennas
that are sensitive to circular polarization.

Simple patch antennas are well-suited to this application because the
narrow bandwidth of a simple patch is an advantage when the signal also has
a narrow bandwidth.  Patches are also very robust mechanically---a
significant advantage in this application.  The polarization of a patch can
be changed from linear to circular simply by moving the feed point.  A
single patch is suitable for the broad beam antenna and the narrow beam
antenna can be made with an array of patches. The ratio of the power in the
array to the power in a single patch will depend only on the array factor,
which is easily calculated. The element spacing of the array can be
adjusted to optimize the beam width and the sidelobe levels.

\begin{figure}
\begin{center}\includegraphics[width=170mm, angle=0]{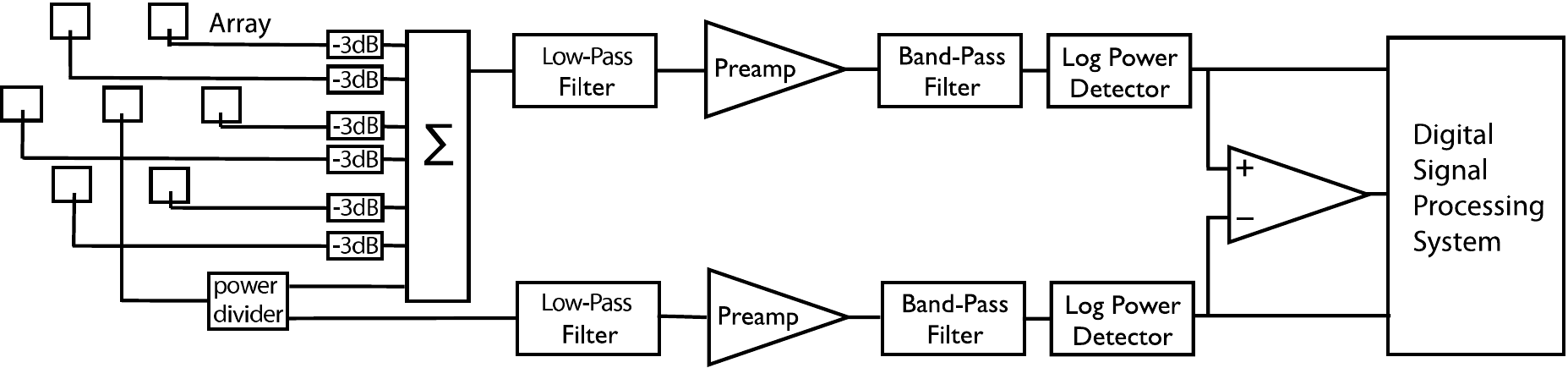}
\end{center}

\caption{Block diagram of the analog signal flow. The patches are shown
with the E field vertical.  The azimuthal angle is defined with respect to
the horizontal. The array is drawn approximately to scale. The center patch
is used both as an array element and as the broad beam element.
\label{fig:block_diagram}}

\end{figure}

The system design consists primarily of: impedance matching a simple patch
antenna at 1090~MHz; adjusting the array configuration to obtain a suitable
beam width and adequately low sidelobes; designing total power detectors
for the two channels; development of signal processing electronics; and
devising a reliable calibration system.  A block diagram of the analog
signal flow is shown in Figure~\ref{fig:block_diagram}. Here one can see
that we have used the center element of the array both as an array element
and as the broad beam element by splitting the signal with a power divider.
To compensate for that power division, the other array elements must have
-3~dB attenuators before the array summer.

\section{Patch Design}

We used a simple square patch excited by a post, fed by coaxial cable from
behind the ground plane. As the desired E polarization is vertical the feed
post must be centered in the horizontal coordinate and offset in the
vertical coordinate, as shown in Figure~\ref{fig:block_diagram}.  Should
circular polarization be desired, the feed point would be moved to the
corner and the horizontal and vertical widths would be made a few percent
different.  The resonant frequency is set by the vertical width of the
patch. The impedance is determined primarily by the feed position and to a
lesser extent by the horizontal width.  We wanted to make the patch using
60 mil dielectric circuit board material, as such a patch has about the
bandwidth we needed.  We tested a patch made with FR4 circuit board
material. We found that FR4 has sufficient dielectric loss at 1090~MHz that
it reduces the Q of the resonance by about 50\%. To avoid broadening the
bandwidth and reducing the gain we used a new low-loss laminate (Rogers
RO4535) which has a loss tangent about 10\% of that of FR4. With this
material the dielectric loss was small compared with the radiated power.

\begin{figure}
\begin{center}\includegraphics[width=140mm, angle=0]{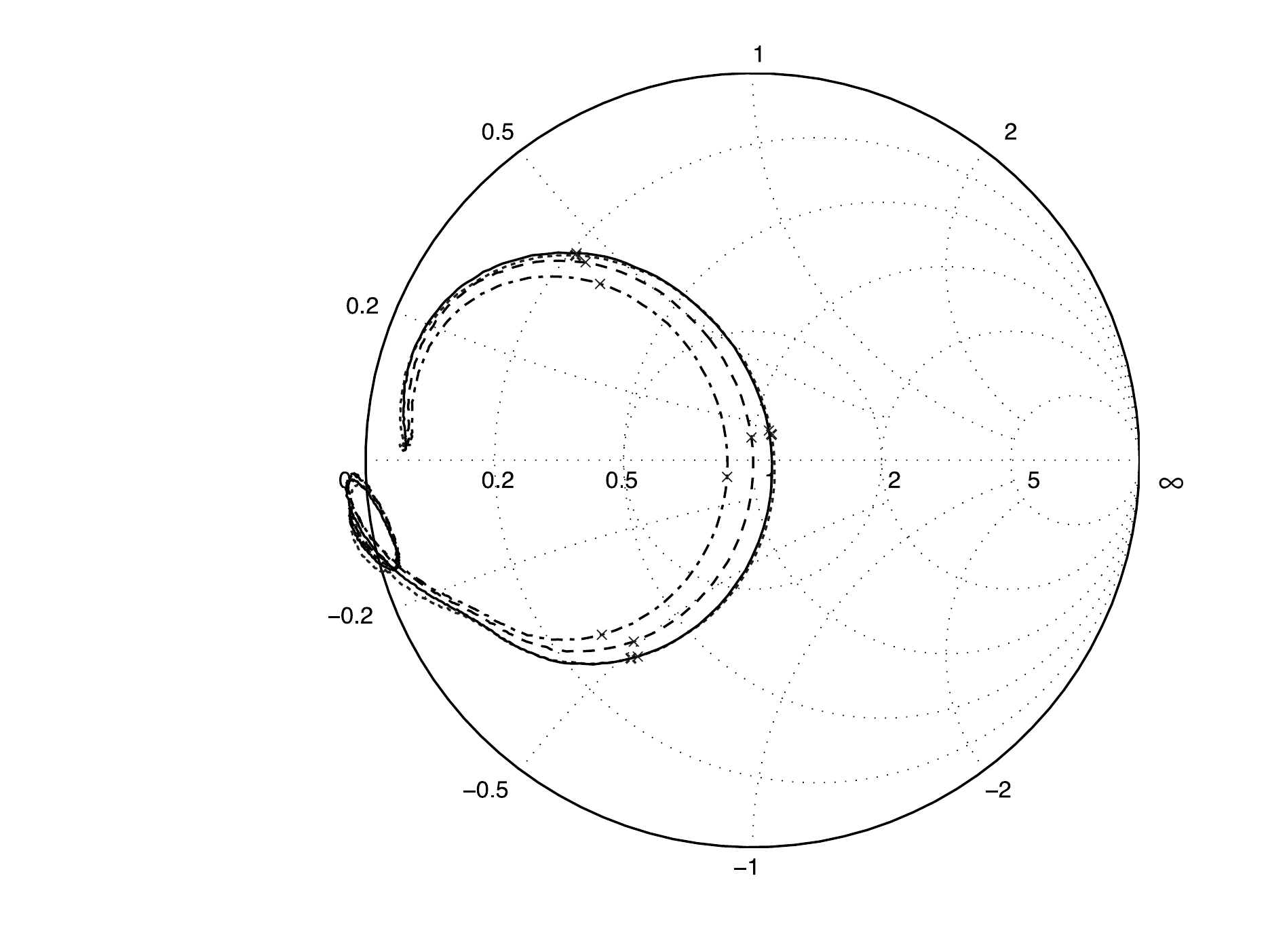}
\end{center}

\caption{The measured driving point impedance of the patch with various
ground planes. The normalizing impedance is 50$\Omega$.  The `x' symbols
are at 1080, 1090 and 1100~MHz. The inner dash-dot trace is the patch
antenna as fabricated, for which the ground plane and dielectric extends
17~mm around the patch. The dashed trace is for the antenna mounted on an
aluminum plate that extends a further 17~mm around the antenna exterior.
The dotted trace is for an aluminum plate that extends 51~mm around the
antenna, quadrupling the original ground plane extent.  The solid trace,
largely overlapping the dotted curve, is for a large plate used to mount
the entire array.  \label{fig:smith_chart}}

\end{figure}

This is a very simple patch design so we did not use an electromagnetic
simulator. We calculated the gain and impedance by modeling the patch as a
pair of slots over an infinite ground plane connected by a microstrip
transmission line with a lossy dielectric. We had to trim the final design
by a few \% after fabrication. The final dimensions were $W_x = 71.0$~mm,
$W_y = 70.2$~mm, and the feed point was 23.3~mm above the lower edge.  The
ground plane occupied the entire back surface of the 105~mm square
dielectric, so that the ground plane fabricated onto the antenna extended
approximately 17~mm beyond the patch boundaries.   The calculated gain for
an infinite ground plane was 5.3~dBi, which includes a dielectric loss of
0.5~dB. The measured gain for a patch mounted on a large ground plane was
$4.7\pm 0.5$~dBi, a reasonable agreement. The impedance was measured for
four different ground planes and is shown in Figure~\ref{fig:smith_chart}.
A ground plane extending 51~mm past the antenna (a total of 68 mm
beyond the patch) is indistinguishable from a much larger ground plane.

The extent of the ground plane is an important factor as it will often be
convenient to minimize the weight and windage of the ground plane material.
In the presence of the full array ground plane there is essentially no back
radiation. However with a reduced ground plane there will be some back
radiation and some change in impedance.  Accordingly we measured the
impedance and the front-to-back ratio of a single patch with different
ground plane widths. The front to back ratio of the three smaller ground planes
shown in Figure~\ref{fig:smith_chart} above, were 9~dB, 13~dB and 22~dB. 
On the basis of these measurements we decided that a 50~mm aluminum
border around the antenna was adequate.

The beam of the patch $G_P(\theta,\phi)$ is relatively broad. An H-plane
cut through $G_P(\theta,\phi)$ has $\sin^2(\theta)$ behavior as shown in
Figure~\ref{fig:gains} with a solid line. The pattern is broader in the E-plane
(shown dashed in Figure 1), dropping
only 2.4~dB at $90^\circ$. We did not attempt to measure the beam width
because a measurement without an adequate test range would be no better
than our calculation, and the exact pattern of the patch is unimportant to
our design because it factors out of the \emph{ratio} of the array gain to
the patch gain.

\section{Array Design}

The array must have a beam width of about $30^\circ$ and sidelobes at
least 10~dB below the main beam. The array will be mounted with its normal
aligned to the optical axis and the array elements will all be equally
phased so the beam is normal to the array plane. The purpose of the array
is to create a narrow beam, which can be compared with the broader beam of
a single element, to determine the angular separation of the transponder
from the telescope optical axis. Thus we are not concerned with the gain of
the array as much as the ratio of its gain to that of a single patch ($R$).
The gain of the single patch factors out of this ratio, simplifying the
calculation. The design problem then is to adjust the patch configuration
to obtain an optimal ratio $R$.

The gain of the array $G_A(\theta, \phi) = C\times G_P(\theta,\phi)\times
AF(\theta,\phi)$. The constant $C$ is determined such that the integral of
$G_A$ over all space = $4\pi$.  The array factor $AF$ is given by

$$AF(\theta,\phi) = \sum_{i=1}^n W_i\exp(-j\mathbf{k}(\theta,\phi)\cdot \mathbf{B}_i) $$

Here $\mathbf{B}_i$ is the vector location of the $i^\mathrm{th}$ patch,
$W_i$ is the (complex) excitation of the $i^\mathrm{th}$ element, and
$\mathbf{k}(\theta,\phi)$ is the wave vector for a plane wave arriving at
the array from the direction defined by $(\theta,\phi)$.  The desired ratio
$R = C\times AF$. Of course one must also allow for the gains of the
amplifiers, losses in the cables and summing junction, etc.

\begin{figure}
\begin{center}\includegraphics[width=89mm, angle=0]{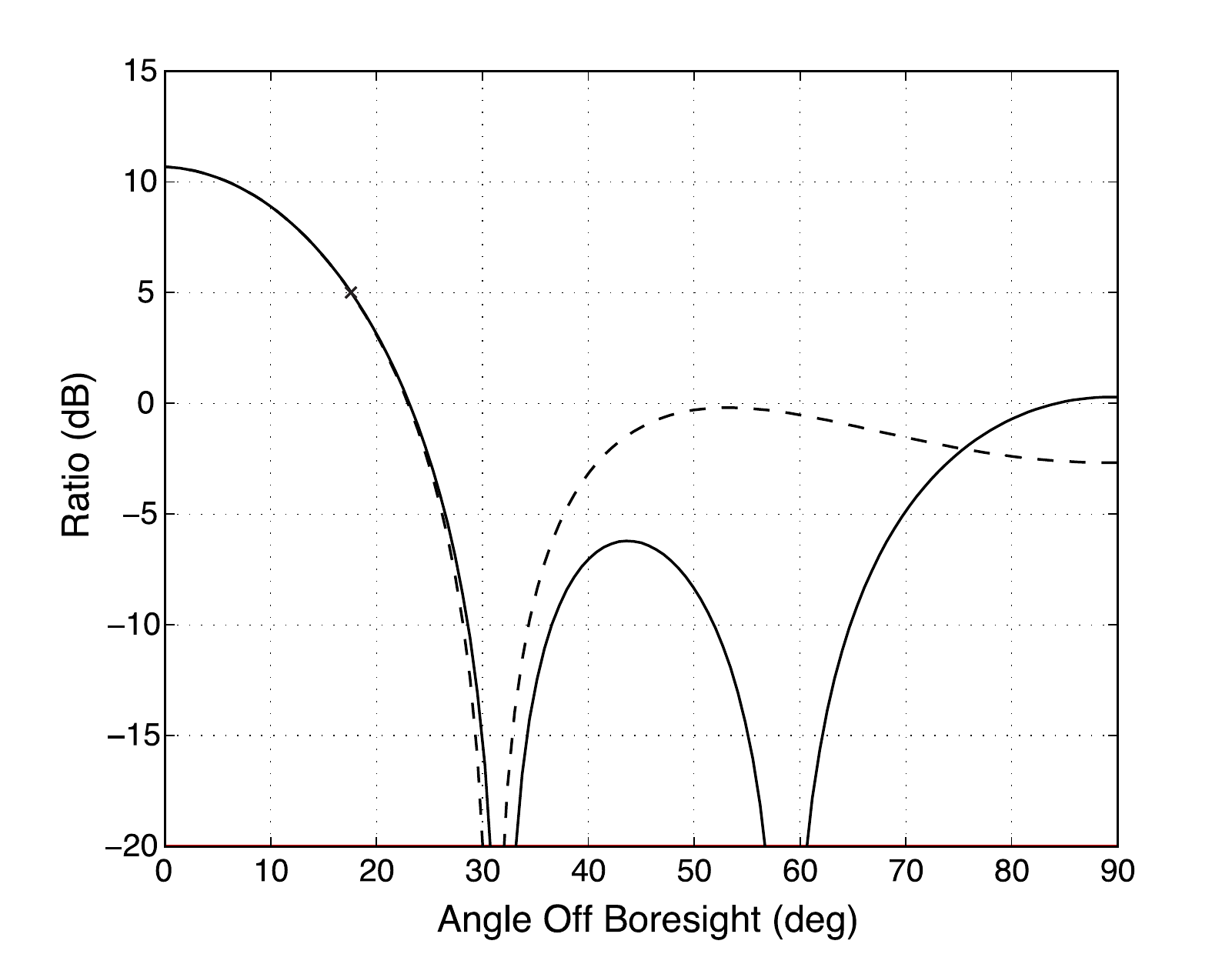}

\end{center}

\caption{Cuts through $R = C\times AF(\theta,\phi)$ at azimuths $\phi =
0^\circ$ (solid line) and $90^\circ$ (dashed line), corresponding
to the highest sidelobes.  The `x' marks a robust detection threshold for
concluding that aircraft is too close to the beam axis.\label{fig:ratio}}
\end{figure}

We searched the space of symmetrical equally-weighted arrays ($W_i = 1$)
with $n$ = 6, 7 and 9 to find a suitable $AF(\theta,\phi)$ with the minimum
$n$. We found a good fit to our requirements with a hexagonal configuration
of $n$ = 7 elements. The pattern has a six part symmetry in azimuth with
sidelobes at two phases. We have defined zero azimuth to be the horizontal as
shown in Figure~\ref{fig:block_diagram}. The sidelobes have equal amplitude
when the element spacing = 0.82$\lambda$. The peak-to-sidelobe ratio is
11~dB. Cuts through the ratio $R$ at azimuths of 0$^\circ$ (the H plane) and
$90^\circ$ (the E plane), which correspond to the highest sidelobes, are shown in
Figure~\ref{fig:ratio}. A detection threshold $R > 5.5$~dB provides the
greatest error margin. This corresponds to an angle of $17^\circ$ off the
axis---close to the design goal and somewhat conservative.

The patch antennas were mounted on an aluminum plate to provide accurate
location, and the beam former was mounted on the same plate. The plate
provides a ground plane, but it is not essential that the ground plane be
continuous provided that the ground extends 50~mm past each antenna. The
outer boundary of the mounting plate was trimmed to provide just the
necessary 50~mm of extent. Slots were also cut within the mounting plate
where possible, to reduce the windage.

We did not expect much mutual coupling with the spacing of 0.82
wavelengths, but we measured the magnitude of $S_{ij}$ between the elements.
We found it was $-27$~dB between colinear elements and $-31$~dB between
diagonal elements. We also measured the driving point impedance of the
center element with all other elements open and also with all other
elements loaded. We could not detect any deviation from the impedance of an
isolated element with a ground plane more than 50~mm around the element.

We did not attempt to measure the pattern of the array because, with
negligible mutual coupling, the calculation is more accurate than any
measurement we could make.  However we did measure the gain at the beam
center. The calculated array gain is 16.3~dBi and the measurement was
$15.95\pm 0.5$~dBi, quite a reasonable agreement.

\section{Receiver Design}

The APOLLO laser is never operated at an elevation less than $15^\circ$
because the accuracy of the lunar ranging is degraded at low elevations by
reduced throughput and atmospheric fluctuations. Thus the greatest range
one can expect for aircraft flying 15~km above the site (itself at 2.8~km
altitude) is 60~km. The gain of a single patch is about 4.7~dB so the
signal level from the minimum-power transponder will be at least $-70$~dBm.
At a distance of 1~km the signal level would increase to about $-35$~dBm.
At the minimum reasonable distance of 100 m, the signal level would be
about $-15$~dBm.

The signal level is so high that the noise figure of the preamp is not a
factor, but interference from the cellular communications band near 900~MHz
can be very strong.  Although the patch antenna has a $-3$~dB bandwidth of
$\approx$1\%, it does not reject a signal at 900~MHz more than about $-20$~dB.
Thus a high-Q multipole band-pass filter is required. Fortunately suitable
filters are available off-the-shelf. We used a 5 pole ceramic filter
\#930644 from the International Microwave Corporation. We also used a sharp-cutoff
low-pass filter to reduce the interference from various communications
above 2~GHz.  Because the signal level is so high, we split the signal from
the center array element and used it both for the array and for the
broad-beam single element. To compensate for the 3~dB loss we inserted 3~dB
attenuators in the path from the other elements to the beam former, as
pictured in Figure~\ref{fig:block_diagram}.

There are a number of inexpensive log amplitude detectors with
good accuracy and wide dynamic range. These are particularly attractive
because the log of the desired power ratio $R$ can be obtained simply using
a differential amplifier to subtract the log of the broad beam power from
the log of the narrow beam power. We used an ZX47-60+ from Minicircuits Lab
with a dynamic range of 5~dBm to $-60$~dBm. To match this to the expected
signal levels we used a  preamplifier with a gain of about 38~dB before the
power detector. The only requirement on the preamp is that it have
relatively low bias current  because it is important to minimize the power
dissipated near the optical axis of the telescope to avoid image
degradation arising from air turbulence. The output of the detectors need 
only pass the narrowest transponder
pulse of 450~ns. We used a 3rd order Butterworth low pass filter
with a 100~ns time constant to minimize the post-detection noise.

\section{Signal Processing}\label{sec:sig_proc}

The signal processing unit performs the dual function of deciding when to
close the laser shutter and capturing the pulse code for logging aircraft
identities and altitudes.  An example pulse pattern is shown in
Figure~\ref{fig:pulses}.  The decision to close the shutter is based on
four criteria, any of which---when satisfied---will result in laser
closure:

\begin{figure}
\begin{center}\includegraphics[width=89mm, angle=0]{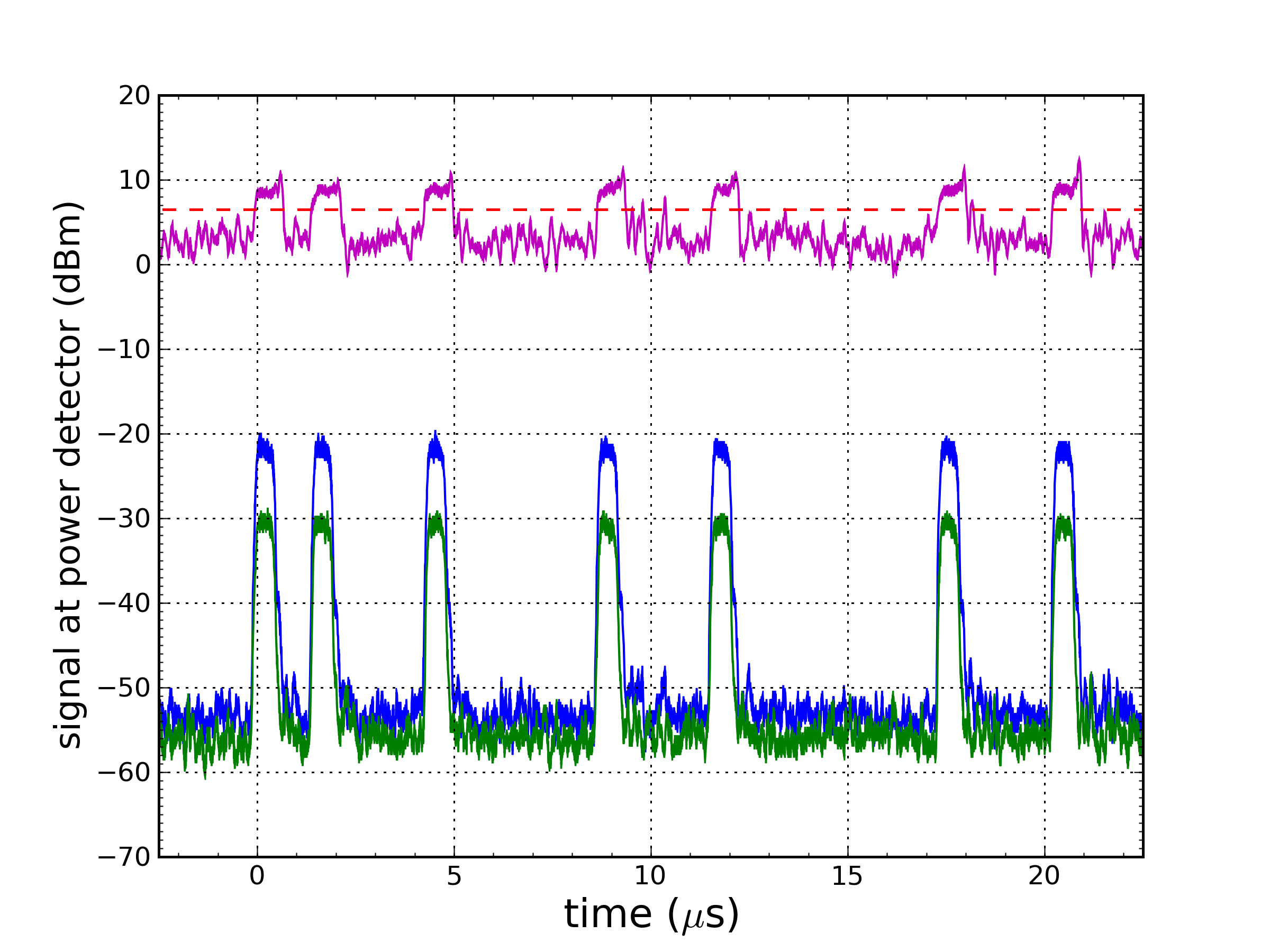}
\end{center}

\caption{Example pulse train obtained in San Diego, showing both the broad
antenna signal (green) and the directional signal (blue), as well as the
output of the difference amplifier (magenta). The two antenna signals are
represented as post-amplifier power at the power detector, while the
difference signal is represented as the ratio, in dB.  The pulse pattern
decodes to 4530, which could either be a Mode-A identity or a Mode-C
altitude (corresponding to an altitude of 3400 feet).  In this case, the
directional signal is stronger than the broad signal, indicating that the
source is within the primary beam of the directional array. The red dashed
line shows where one might place a threshold for judging ``in beam''
activity.  The trailing edge spikes are not uncommon, and motivate using
only the first part of the pulse for discrimination, as discussed in the
text.\label{fig:pulses}}

\end{figure}

\begin{enumerate}
\item D1 = DIREC/BROAD ratio $R > 5.5$~dB, AND DIREC signal $> -24$~dBm at detector
\item D2 = DIREC signal $> -4$~dBm at detector
\item D3 = BROAD signal $> -4$~dBm at detector
\item D4 = $|\mathrm{Power\ supply\ current - Nominal} | > 0.05 \times$ Nominal
\end{enumerate}

The first criterion is the most important, representing the
directionally-sensitive detection mode.  The second criterion avoids a
failure mode of the first criterion due to saturation of the DIREC signal
from a nearby source The third criterion prevents a nearby,
fast-angular-motion airplane from getting into the protected zone before
the other two criteria are activated---keeping in mind that we rely on
external interrogation occurring before sensing the presence of the
aircraft.  Each of these decisions is based on comparators sensing the
outputs of the logarithmic power detectors and referenced to an adjustable
voltage, with the DIREC/BROAD ratio provided by a difference of the
logarithmic outputs.  The angular size of the protected zone on the sky is
adjustable by setting the difference-comparator reference voltage.

\begin{figure}
\begin{center}\includegraphics[width=89mm, angle=0]{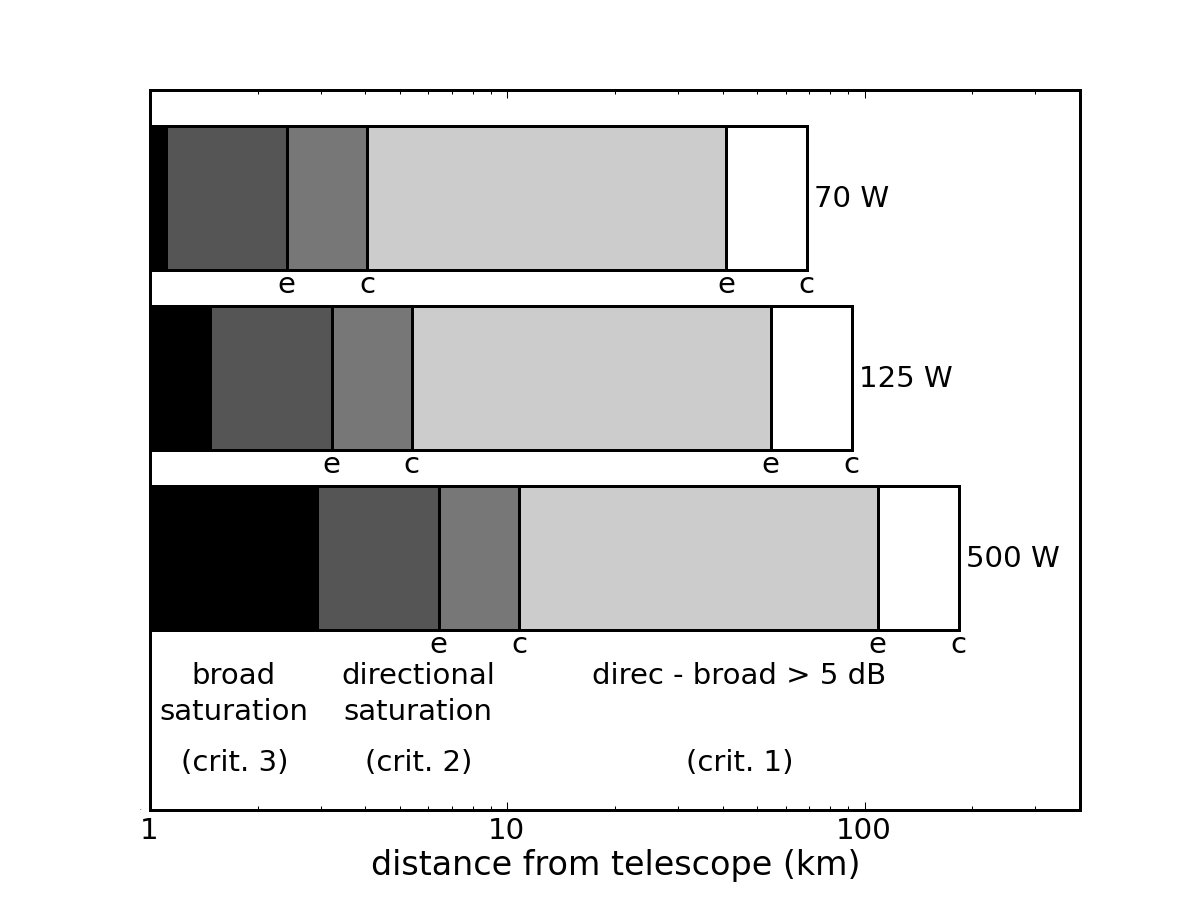}
\end{center}

\caption{Protected line-of-sight ranges for the first three shutter closure
criteria described in the text for three different transmitter powers
corresponding to: 70~W---the minimum for non-commercial (typically low
altitude) aircraft; 125~W---the minimum for commercial aircraft; and
500~W---the maximum power permitted.  For each, the zone for which the
broad channel saturates is black; the zone for which the directional
channel saturates is medium gray; and the zone in which the ratio criterion
is relied upon is a light shade.  For the latter two, different shadings
represent response at the edge and center of the 15-degree half-width
protected area on the sky, denoted by ``e'' or ``c'', respectively.
\label{fig:protect}}

\end{figure}

The criteria based on signal levels map into different nominal ranges for
the three different peak power requirements imposed by the FAA.  For
non-commercial traffic (typically low-altitude), the minimum peak power
requirement is 70~W.  For commercial aircraft, the minimum requirement is
125~W.  The maximum allowable peak power is 500~W.
Figure~\ref{fig:protect} illustrates the distances to which the first three
criteria in the list above apply for the three limiting cases of
transmitted power.  For example, an airplane with a 125~W transmitter will
saturate the broad antenna channel (criterion 3) if within 1.9 km, will
saturate the directional channel (criterion 2) within 3.2~km to 5.4~km
depending on where it is within the directional beam, and will be protected
by the ``in beam'' criterion (the first in the list) out to 55 to 92~km
depending again on beam position.

Any of the three triggers activates a re-triggerable one-shot, holding the
laser shutter in a closed condition for five seconds.  Reaction time is
within 30~$\mu$s of signal receipt.  The system is not confused by multiple
aircraft in the beam at the same time.  Any signal deemed to meet the above
criteria will activate the shutter.  The signal processing unit also
monitors the power supply current to ensure that the components are
operating normally before allowing the laser shutter to open.

The signal processing unit also captures the altitude and identity codes
transmitted from aircraft that have triggered shutter closure, for the
purpose of recording activity.  This is done using a PIC microprocessor
responding to the initial pulse via an interrupt service routine, then
checking at regular intervals (about 1.45~$\mu$s) for any pulse transitions
within the previous pulse period.  Though coded to interpret Mode-A
(identity) and Mode-C (altitude) responses, the characteristic pattern of
Mode-S transmissions can be discerned, and distance measuring equipment
(DME) signals operating on the same transmit frequency band can also be
identified.  Note that any signal meeting the criterion established above
will close the laser shutter regardless of information content.  The
information is used to build a data base of the frequency and nature of the
aircraft-caused triggers---including a crude estimate of distance, given
telescope elevation angle and aircraft altitude.  The microprocessor can
also assume control of the shutter and build in false-trigger avoidance.
Electrostatic discharge or lightning generally produces a single pulse,
while aircraft signals (Mode-A, Mode-C, Mode-S, DME) all contain multiple
pulses within 20~$\mu$s.  These ``glitches" may be ignored.  

We have found that multi-path interference---especially seen by the broad beam
antenna---often compromises the pulse quality, and often produces false
triggers.  We have successfully tested a method to mitigate such false
triggers, by applying criterion 1 above only during a 50~ns window shortly
following the leading edge of a clean pulse (one that has only background
noise preceding it).  We have seen this approach dramatically reduce the number
of false ``in-beam" triggers while not compromising the robust detection
of true in-beam events.

Besides the check that the current delivered to the electronics is good,
the microprocessor sends a keep-alive packet once per minute to assure
system health.  It is straightforward to verify that the system is
operational before commencing with laser activities.  Logged information is
sufficient to verify sensitivity to aircraft on a nightly basis, but an
in-dome calibrated transmitter may also be useful to verify proper
sensitivity.

\begin{figure}
\begin{center}\includegraphics[width=89mm, angle=0]{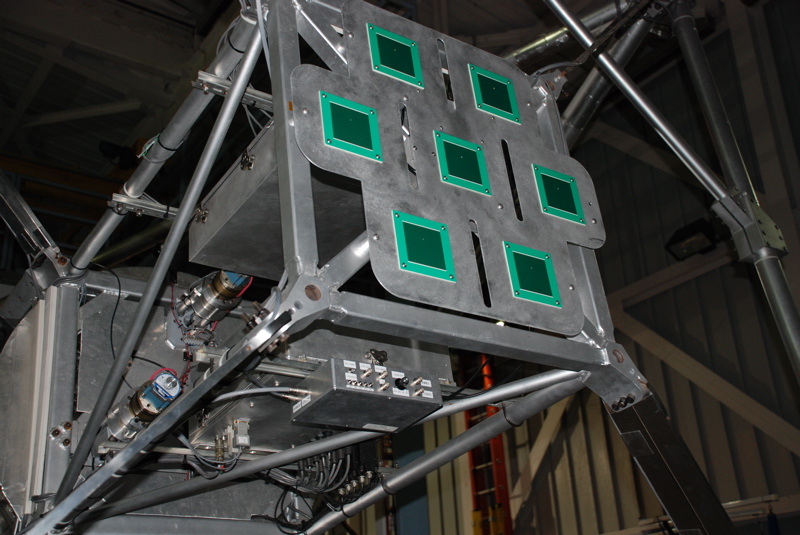}
\end{center}

\caption{The antenna array mounted on the sky-facing side of the secondary
mirror support structure on the APO 3.5~m telescope.  The electronics box
with white labels visible below the antenna plate contains the RF
electronics, and consumes only 3~W of power.\label{fig:onscope}}

\end{figure}

The system is implemented as three physical units, all designed and built
at the University of California, San Diego (UCSD).  The antenna array and
passive summing devices are mounted to a single aluminum plate 0.63~m
across that also serves as an extended ground plane.  The RF processing and
discrimination is performed in a small electronics enclosure measuring
$22\times 15\times 6$~cm$^3$ that is located close to the antenna.  These
electronics only consume 3~W of power, so that the box may be safely
deployed in front of a telescope without disrupting image performance.  The
microprocessor and power supply occupy a second box measuring $19\times
12\times 8$~cm$^3$ that may be located tens of meters from the antenna RF
electronics.  A connection from the microcontroller box to a computer or
terminal server completes the system.  Figure~\ref{fig:onscope} shows the
antenna mounted on the 3.5~m telescope at APO, on the sky-facing side of
the secondary mirror support structure.

%\section{Calibration}
%
%The ratio signal $R(\theta,\phi,t)$ is pulse width modulated. The pulses are
%either positive (if the aircraft is near the beam axis) or negative
%(otherwise) and the off-pulse level is the difference between the noise
%level of the two logamp detectors (close to zero). The amplitude of the
%maximum positive pulse $R_\mathrm{MAX}$ is the primary reference. It should
%be determined by measurement because the gains in the narrow beam and broad
%beam receiver channels are not perfectly matched and this causes an offset
%from the calculated value shown in Figure~\ref{fig:smith_chart}. We
%determined $R_\mathrm{MAX}$ simply by pointing the system at the stream of
%aircraft in the landing pattern at San Diego International Airport.

\section{Performance}

\begin{figure}
\begin{center}\includegraphics[width=89mm, angle=0]{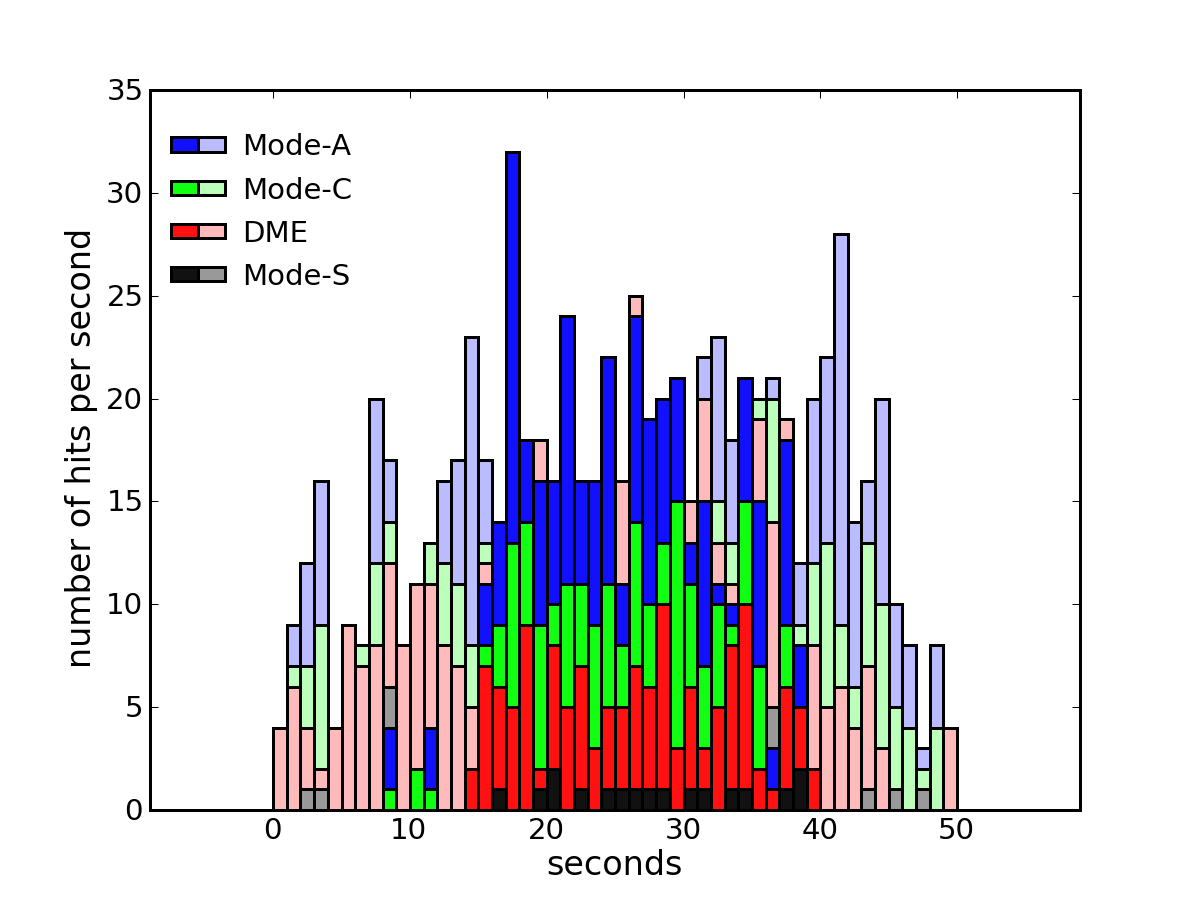}
\end{center}

\caption{Event rate in 1~s bins for a pass acquired on 31 December 2008,
for an airplane squawking identity code 6755 at an altitude of 39,000~ft,
while the telescope was at an elevation angle of 54$^\circ$.  Event types
are coded as blue for identity (Mode-A), green for altitude (Mode-C), red
for DME, and black for Mode-S.  Saturated shades represent those detections
deemed to be in the central beam by criterion 1 in
Section~\ref{sec:sig_proc}.\label{fig:pass}}

\end{figure}

The antenna system was deployed on the Apache Point Observatory 3.5~m
telescope on 2008 December 19 for initial characterization.  Besides the
trigger criteria detailed above, an alternate scheme is employed for the
purpose of recording a greater amount of aircraft traffic as detected in
the broad-beam antenna.  Setting the detection threshold to sense airplanes
within 20~km at 70~W peak power, or 52~km at 500~W, we typically record
about 6 airplanes (maximum of 20) per night through the slit of the
telescope enclosure, when open.  This is based on 72 full nights of
operation between 2008 December 31 and 2009 August 7.  Note that the solid
angle of visible sky is restricted by the enclosure to a range of 1.9--2.7
steradians depending on telescope elevation angle, averaging only 36\% of
the sky.  On a nightly basis, about 2--3 (maximum of 9) airplanes cross the
threshold to qualify as an ``in-beam" detection, resulting in a median of
70~s (maximum 283~s) of shuttered time per night, out of about 42,000~s of
open time ($\sim$0.2\% closure).  As yet, all closure events are associated
with real aircraft.  Typical detection rates are about 12 events per second
during a beam-crossing pass, about 40\% of which are associated with Mode-A
identity codes, 35\% associated with Mode-C altitude codes, 20\% with
distance measuring equipment (DME) pulses, and 5\% identified as Mode-S
information packets.  We decode and record the Mode-A and Mode-C
information, but do not decode the Mode-S information with the present
firmware.  Figure~\ref{fig:pass} demonstrates the behavior of a typical
beam-crossing detection.  The metal telescope enclosure shields the antenna
from line-of-sight detection at large angles, which results in a relatively
tight truncation of the sequence.  The central beam crossing is robustly
detected in 408 events, roughly centered in the crossing of the open
enclosure slit.

Tests of a second, improved system at the UCSD campus under heavy
air traffic revealed a susceptibility to false triggers due to multi-path
interference and coherent interference from overlapping pulses.  When the antenna was pointed 25$^\circ$ off the horizon
toward the San Diego International Airport 17~km to the south, the shutter
was closed 40\% of the time---and 80\% during the busiest hours of the day.
Implementing a 50~ns window in which to make the ``in-beam" decision just
after the leading edge of the individual pulses reduced this to 7\% of the
time.  Additional filtering by the microprocessor to accept only events
deemed to be in the beam more often than 10 times in any 5 second interval
reduced many of the remaining spurious triggers so that the shutter was
closed 3\% of the time.  More drastic filtering revealed a hard ``floor" at
2\% of the time, suggesting this to be the actual fraction of time aircraft
occupied our beam.  Zenith pointings at UCSD resulted in shutter closures
3\% of the time without the leading-edge window filter, 0.4\% with the
filter, and 0.1\% of the time with the additional software filtering
activated.  Observatories tend to be in far less congested areas, so that
these interference issues---while manageable---will be less important.

\acknowledgments

The authors acknowledge the donation of a new low-loss laminate RO4535 for
the construction of the patches from the Rogers Corporation. We also
acknowledge the contribution of Mike Rezin who assisted in producing the
patch antennas and electronics assemblies.

\end{document}